\documentclass[journal=ancac3,manuscript=article]{achemso}

\usepackage{chemformula} % Formula subscripts using \ch{}
\usepackage[T1]{fontenc} % Use modern font encodings

\author{Pavel V. Kolesnichenko}
\affiliation[Swinburne University of Technology]{Centre for Quantum and Optical Science, Swinburne University of Technology, Melbourne, Victoria 3122, Australia}
\altaffiliation{ARC Centre of Excellence in Future Low-Energy Electronics Technologies, Swinburne University of Technology, Melbourne, Victoria 3122, Australia}

\author{Qianhui Zhang}
\affiliation[Monash University]{Monash University, Melbourne, Victoria 3800, Australia}

\author{Changxi Zheng}
\affiliation[Monash University]{Monash University, Melbourne, Victoria 3800, Australia}
\altaffiliation{ARC Centre of Excellence in Future Low-Energy Electronics Technologies, Monash University, Victoria 3800 Australia}

\author{Michael S. Fuhrer}
\affiliation[Monash University]{Monash University, Melbourne, Victoria 3800, Australia}
\altaffiliation{ARC Centre of Excellence in Future Low-Energy Electronics Technologies, Monash University, Victoria 3800 Australia}

\author{Jeffrey A. Davis}
\affiliation[Swinburne University of Technology]{Centre for Quantum and Optical Science, Swinburne University of Technology, Melbourne, Victoria 3122, Australia}
\altaffiliation{ARC Centre of Excellence in Future Low-Energy Electronics Technologies, Swinburne University of Technology, Melbourne, Victoria 3122, Australia}
\email{jdavis@swin.edu.au}

\title[Localized charge dendrites]
  {Disentangling the effects of doping, strain and defects in monolayer WS$_2$ by optical spectroscopy}

\begin{document}

\begin{tocentry}

\includegraphics[width=\linewidth,height=\textheight,keepaspectratio]{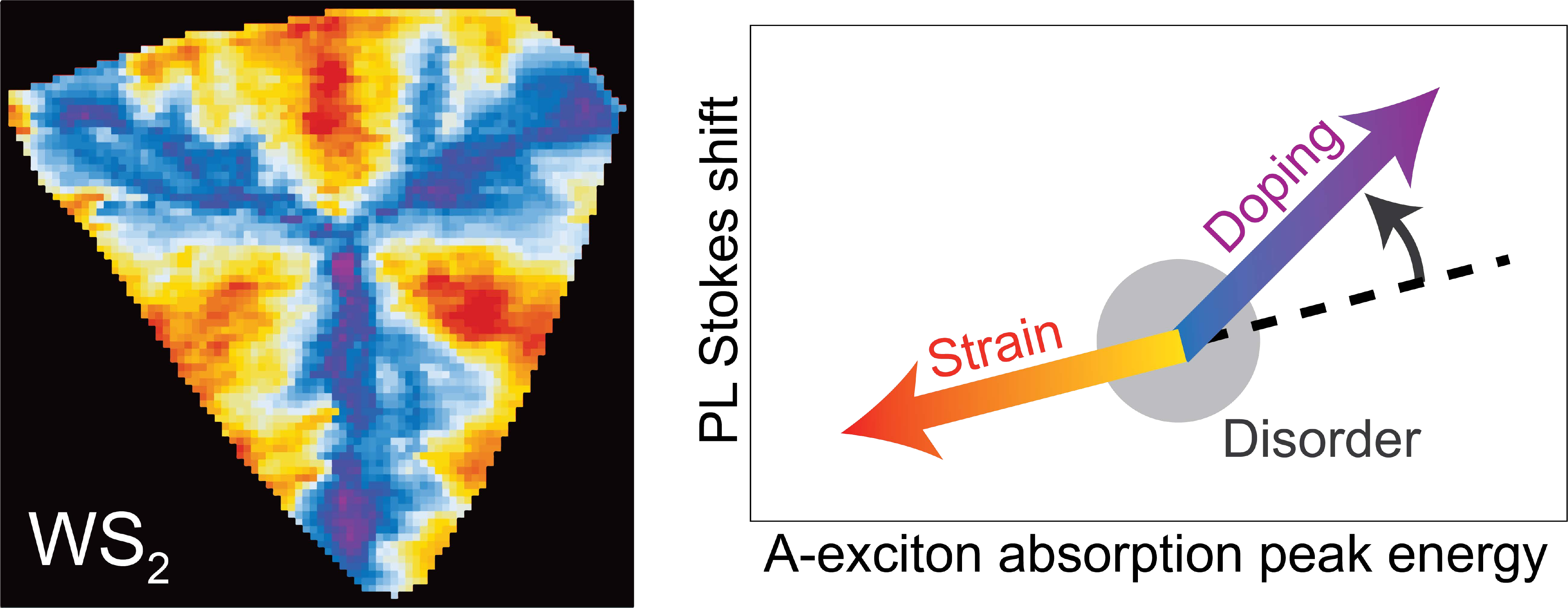}

\end{tocentry}

\begin{abstract}
Monolayers of transition metal dichalcogenides (TMdC) are promising candidates for realization of a new generation of optoelectronic devices. The optical properties of these two-dimensional materials, however, vary from flake to flake, or even across individual flakes, and change over time, all of which makes control of the optoelectronic properties challenging. There are many different perturbations that can alter the optical properties, including charge doping, defects, strain, oxidation, and water intercalation. Identifying which perturbations are present is usually not straightforward and requires multiple measurements using multiple experimental modalities, which presents barriers when attempting to optimise preparation of these materials. Here, we apply high-resolution photoluminescence and differential reflectance hyperspectral imaging \textit{in situ} to CVD-grown WS$_2$ monolayers. By combining these two optical measurements and using a statistical correlation analysis we are able to disentangle three contributions modulating optoelectronic properties of these materials: electron doping, strain and defects. 
In separating these contributions, we also observe that the B-exciton energy is less sensitive to variations in doping density than A-excitons. 
%We show that this can be attributed to variations in the conduction band spin-orbit splitting. contrasting previously reported observations in doping-dependent studies of MoS$_2$ monolayers. We concluded that monolayers of WS$_2$ in this work feature dendritic patterns of localized charge not reported previously and suggest a non-equilibrium competition-supported kinetic growth process undergoing a fractal-to-compact phase transition.
\end{abstract}

\section{Introduction}

For more than a decade two-dimensional (2D) transition metal dichalcogenides (TMdCs) have been extensively studied as they offer unprecedented physics not achievable in conventional semiconducting quantum wells. The peculiar combination of their crystal structure without inversion symmetry\cite{Xiao2012,Liu2015}; strong spin-orbit (SO) interactions originating from the heavy transition metal elements\cite{Liu2015}; and the time-reversal symmetry acting to lock spin and valley degrees of freedom makes these materials ideal platforms for realization of valley-selective\cite{Schaibley2016} and spin-polarized\cite{2019} optoelectronics. Owing to their monolayer nature, the optoelectronic properties of these materials are governed by excitonic effects that are greatly enhanced compared to conventional semiconducting quantum wells\cite{Wang2018}. Several conceptual devices based on 2D materials, such as field-effect transistor\cite{Georgiou2012}, memory cells\cite{Bertolazzi2013,Choi2013}, \textit{p}-\textit{n} junction\cite{Deng2014} and photodiodes\cite{Furchi2014,Lee2014} have been already demonstrated to have the potential to outperform their conventional analogs.

%While mechanically exfoliated 2D TMdCs tend to exhibit a better crystalline quality compared to those grown by CVD techniques\cite{Hong2015}...

However, the majority of the demonstrated devices are based on "one-off" prototypes based on mechanically exfoliated monolayers\cite{Georgiou2012,Choi2013,Bertolazzi2013,Furchi2014,Lee2014} and  their fabrication lacks scalability. Chemical vapour deposition (CVD) is promising for massive industrialization of novel optoelectronic devices, but, as it currently stands, monolayers grown via CVD are randomly distributed across their supporting substrates with each monolayer flake having heterogeneous optoelectronic properties. These heterogeneities are often challenging to interpret as the crystal structure can be subject to a complex perturbation resulting from  strain\cite{Hsu2017}, defects\cite{Jeong2015,Jeong2017,Lin2017,Rosenberger2018,Kastl2019}, grain boundaries\cite{Zande2013,Zhang2013,Bao2015,Liu2016,Kim2016,Kastl2019} as well as oxidation\cite{Zhang2013}, water intercalation\cite{Zheng2015} and other aging processes\cite{Gao2016}. Despite this variability and complexity, CVD approaches commonly result in triangular monolayers with variations in the optical properties that are also trigonally-symmetric\cite{Cong2013,Peimyoo2013,Bao2015,Jeong2015,Liu2016,Yore2016,Gao2016,Kim2016,McCreary2017,Jeong2017,McCormick2017,Lin2017,Borys2017,Sheng2017,Rosenberger2018,Bogaert2018,Kastl2019}. Most simply, these correspond to regions of brighter (bright regions) and darker (dark regions) photoluminescence (PL) emission. The dark regions have been commonly attributed to intra-flake grain boundaries \cite{Bao2015,Liu2016,Kastl2019}, structural or chemical heterogeneities \cite{Cong2013,Peimyoo2013,Sheng2017,Jeong2017,Lin2017,Bogaert2018}, and/or linked to increases in $n$-doping. Identifying the specific perturbation or combination of them is challenging, and the details will vary depending on the specifics of the growth, substrate, environment and history of the monolayer.

Disentangling various contributions modulating optoelectronic properties of TMdC monolayers usually requires complementary and often complicated methods. These can include optical and Raman spectroscopy, atomic force microscopy (AFM), Kelvin probe force microscopy (KPFM), and photoemission spectroscopy, amongst others.
% transfer of monolayers between substrates,  Further, determination of intrinsic doping levels is usually possible by electrical gating, which requires an additional effort of providing metallic contacts and connecting them to a current/voltage controller. In addition, in places where contacts are introduced, the crystalline quality of monolayer flakes is also altered. Kelvin probe force microscopy (KPFM) is another approach to determine the relative levels of doping, however, often the results are challenging to acquire. Finally, the presence of defects can be determined by various techniques such as X-Ray photoemission spectroscopy (XPS)\cite{Kastl2019}, nano-Auger spectroscopy\cite{Bao2015}, electron energy-loss spectroscopy (EELS)\cite{Lin2017}, conductive atomic force microscopy (AFM)\cite{Rosenberger2018} or decorating TMdC monolayers by silver nanoparticles\cite{Jeong2015,Jeong2017}. 
While all these methods have their own advantages they usually require repositioning of the monolayer from one experimental setting to another. 

Here we demonstrate the ability to disentangle the effects of doping, strain and defects, which play a major role in shaping the optoelectronic properties of TMdC monolayers, using solely optical spectroscopy measurements.
To acheive this we record PL and absorption hyperspectral images of WS$_2$ monolayers grown via CVD with high spatial resolution ($\sim$300--380 nm). %  in the conditions of aggressive nucleation% which resulted in a dendritic geometry of the dark regions. %The dendritic nature of the obtained patterns provides a larger effective surface area and therefore a higher number of spatially-resolved data-points enabling correlation statistical analysis of optoelectronic properties within both bright and dark regions.
Among the countless reports showing spatial maps of PL spectral characteristics of TMdC monolayers, only few presented spatially resolved absorption characteristics\cite{Dhakal2014,Park2016,Kim2016,Castellanos-Gomez2016,Nozaki2016,Borys2017,Yuan2018},  and even fewer with the spatial resolution less than 500~nm\cite{Dhakal2014,Nozaki2016,Castellanos-Gomez2016,Yuan2018}.  Having both emission and absorption hyperspectral maps provides a greater level of detail and allows parameters such as PL Stokes shift and PL quantum yield to be determined, which help to identify the different perturbations. To analyse the large amount of data, we extract different spectral properties (including peak amplitude, wavelength and width for A-excitons, B-excitons, and trions) from each point in space and perform a correlation analysis which ultimately allows us to disentangle the effects of strain, doping and disorder. This idea of correlating different spectral properties to better understand 2D materials has previously been shown to provide useful insights when applied to Raman and PL spectroscopy measurements on graphene\cite{Lee2012}, TMdC monolayers\cite{Bao2015,Michail2016,Hsu2017,Kastl2019} and their heterostructures\cite{Rao2019}. We show here that correlating PL and absorption measurements can provide similar insights into the properties and pertubations of 2D materials.

\section{Results and discussion}

We examine two monolayer WS$_2$ flakes grown by CVD on sapphire with properties that encompass the range of observed behaviours. The results for a further five flakes are shown in the Supporting Information. The PL hyperspectral images were obtained by illuminating the sample with a 410~nm cw laser and collecting the emission with a 100x objective in a modified confocal microscope as detailed in the Experimental section. 
% The PL spectrum was recorded at each position as the sample position was scanned in ****nm steps.  Spatial heterogeneities of WS$_2$ monolayers were investigated by high-resolution PL and DR hyperspectral imaging. Overall seven monolayer flakes (flakes~\#1--7) were characterised. The results for two flakes with properties that encompass the range of observed behaviours are presented in the main text (labelled flakes~\#1 and \#2). The results from the other monolayers are given in Supporting Information.

\begin{figure}[h!]
\centering
\includegraphics[width=\linewidth,height=\textheight,keepaspectratio]{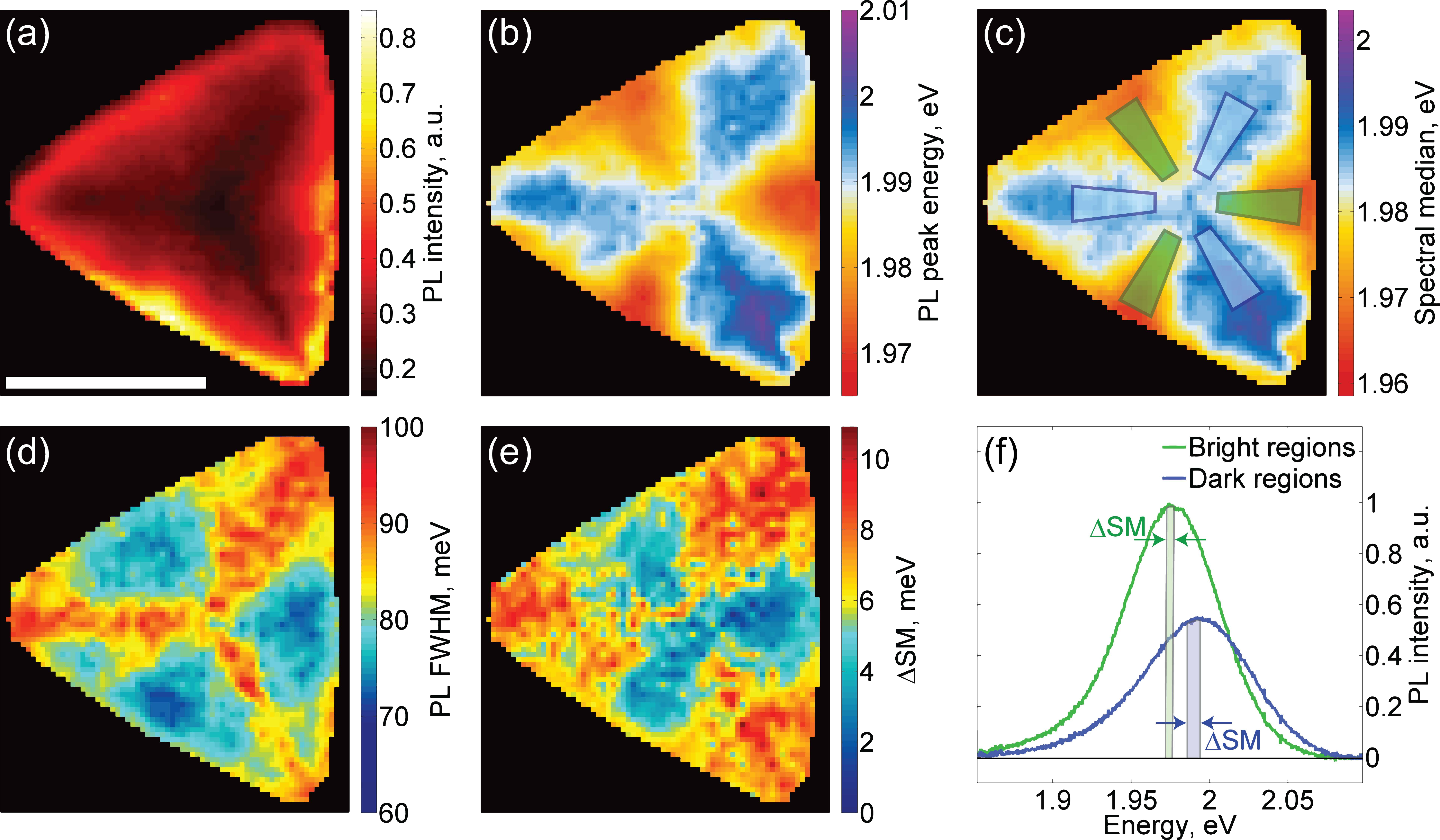}
\caption{Flake~\#1: PL spatial maps of (a) normalized integrated intensity, (b) peak energy, (c) spectral median, (d) FWHM, and (e) $\Delta$SM. (f) Normalized integrated PL spectra across the dark (blue) and bright (green) regions. The spectra are normalized to the peak intensity of PL spectrum integrated across the bright regions. The corresponding integration areas are shown in (c). The length of the scalebar in (a) corresponds to 10 $\mu$m.}
\label{fig:flake2_PL_Fig1}
\end{figure}

Figure~\ref{fig:flake2_PL_Fig1}a maps the integrated PL intensity across flake \#1, revealing a commonly observed trigonally-symmetric pattern. In the interior these are categorised broadly as (i) \textit{dark regions}, and (ii) \textit{bright regions}. At the edges of the flake the brightest PL is observed and defines a third domain. These bright edges have been commonly observed for TMdC monolayers on hydrophilic substrates such as sapphire\cite{Zheng2015}. % and UV-exposed rutile\cite{Paz2011,Kastl2019}. Edges were previously found to contain a much higher density of sulphur vacancies (S-vacancies) compared to the interior\cite{Kim2016,Carozo2017,Rosenberger2018,Kastl2019} due to the rapid termination of the growth process at the cooling stage of a CVD cycle. 
In these regions water intercalation\cite{Zheng2015} and interaction of sulphur (S) vacancies with environmental molecules\cite{Tongay2013,Nan2014,Chow2015,Sun2017} reduces the intrinsic $n$-doping levels commonly found in these materials resulting in a brighter PL emission. %The interaction of edges with growth promoters distributed across the substrate  may also increase PL intensity at the edges of TMdC monolayers\cite{Kim2016}. 
Here, we focus on the interior of WS$_2$ monolayers and will omit further discussion of edges.

The integrated and normalized PL spectra from the dark  and bright regions are shown in Figure~\ref{fig:flake2_PL_Fig1}f. The asymmetric spectral shape of the PL spectra is typical for CVD-grown WS$_2$ monolayers and arises from the dominant emission from the A exciton, with the low energy tail coming from trions and/or defect states. To further investigate the spatial variation across the flake, PL peak energy (Figure~\ref{fig:flake2_PL_Fig1}b), spectral median (SM) (Figure~\ref{fig:flake2_PL_Fig1}c), and PL FWHM (Figure~\ref{fig:flake2_PL_Fig1}d) were mapped across the sample. These spatial maps show that the PL emission from the dark regions is blue-shifted (Figure~\ref{fig:flake2_PL_Fig1}b) and broadened (Figure~\ref{fig:flake2_PL_Fig1}d) compared to that from the bright regions. The SM is similar to the peak wavelength and is defined as the energy that splits the PL spectral profile into two equal contributions such that the total spectral weight to the left of the SM is equal to the spectral weight to the right of the SM. In contrast to the peak energy, however, the SM is sensitive to changes in the asymmetry of the peak and thus changes to the trion contribution. To highlight the difference, we also plot $\Delta$SM - the difference between the SM and the peak energy (Figure~\ref{fig:flake2_PL_Fig1}e). The value of $\Delta$SM (Figure~\ref{fig:flake2_PL_Fig1}e) is predominantly defined by the charging energy (CE) of a trion (see Supporting Information) and represents a good, fit-free, indication of how the CE varies across the flake. As seen from Figure \ref{fig:flake2_PL_Fig1}e,f, the value of $\Delta$SM (and, therefore, CE) is larger in the dark regions, which suggests that the doping density may be higher in these regions, which would also lead to the reduced emission intensity observed\cite{Mak2012}. %Indeed, increased doping has previously been suggested as the origin for reduced emission in the dark regions\cite{}. 
However, there are also other mechanisms that could explain the variations of the PL emission properties across the flake. For example, the weaker, blue-shifted, and broadened PL in the dark regions could originate from intra-flake grain boundaries\cite{Bao2015,Liu2016,Barja2016,Kastl2019}.
%These grain boundaries may introduce metallic states\cite{Zande2013,Zou2015,Ma2017} reducing the overall PL emission. The presence of grain boundaries also implies that the overall energy of the crystal is not minimized in those localities contributing to an observed energy difference between the dark and bright regions. This may induce diffusion of excitons via funneling\cite{Feng2012} and interaction with optical phonons, facilitating the migration of excitons from the higher-energy dark regions towards the lower-energy bright regions. This diffusion could be an additional reason of lower PL emission in the dark domains. Another contribution shaping optoelectronic properties within the dark regions could be from increased doping levels in the dark regions which was shown to lead to a blue-shifted, broadened and dimmer PL emission in 2D semiconductors\cite{Schmitt-Rink1989}.
%In contrast to the dark regions, bright regions emitting red-shifted and narrowed PL could be considered as the domains corresponding to a higher crystal quality. However, brighter and narrower PL emission in ambient conditions does not necessarily imply higher crystalline quality of 2D TMdCs\cite{Tongay2013}, and various mechanisms can enhance, narrow and/or red-shift the emitted PL. First, 
On the other hand, the increase in intensity and red-shift in the bright regions could be due to environmental molecules interacting with S-vacancies, which is known to increase the PL quantum yield (QY) \cite{Tongay2013,Nan2014,Chow2015,Sun2017}. Tensile strain has also been shown to lower emission energy\cite{Feng2012,Castellanos-Gomez2013,Zhang2016,Frisenda2017} and alter the exciton-phonon coupling with the effect of narrowing the A-exciton linewidth\cite{Khatibi2018,Niehues2018}. The presence of tensile strain in the bright regions has been reported previously for CVD-grown monolayers with similar trigonal-symmetric patterns\cite{McCreary2016,McCreary2017}.% It was shown that following the transfer of the monolayer to another substrate, the PL intensity from the bright regions was significantly reduced with increased emission energy\cite{McCreary2016} suggesting relaxed tensile strain field in these regions. In addition, bright regions were shown to exhibit relaxed valley optical polarization selection rules\cite{McCreary2017} which may reflect the broken \textit{C}$_3$-rotational symmetry due to strain or defects. Finally, it was suggested that for W-based monolayers the presence of a tensile strain field  matching our observations within the bright domains.

From these PL maps it is evident that the integrated intensity, peak energy, FWHM and CE appear to be strongly correlated and vary together across the flake. These variations could be due to gradually changing strain, defect density, doping density or grain boundaries. In order to differentiate between these possibilities, we correlate these PL maps with absorption maps. 
%It is clear that there are several possible causes of the variations in the PL properties across the flake and from solely PL measurements it is not possible to determine the physical origins of the observed patterns: a complementary measurement is, therefore, required. In this work we perform spatially-resolved absorption measurements that can provide additional details and facilitate the interpretation of the observed patterns. In particular, in addition to the properties of A-excitons, absorption measurements can provide information on other types of excitons typical to this family of monolayers such as B- and C-excitons. Moreover, correlating spatial maps of the PL emission spectral characteristics with those reflecting absorption properties can provide correlated PL-absorption physical quantities such as PL Stokes shift and relative PL QY.

%\begin{figure}[h!]
%\centering
%\includegraphics[width=0.5\linewidth,height=\textheight,keepaspectratio]{flake2_DR_Fig2}
%\caption{Flake~\#1: DR spatial maps of (a) peak amplitude and (b) peak energy of A-exciton transition. (c) Integrated DR spectra across the dark dendritic (blue) and bright (green) regions. The spectra are normalized to the levels of B-exciton spectral peak. The corresponding integration areas are shown in (b). The length of the scalebar in (a) corresponds to 10 $\mu$m.}
%\label{fig:flake2_DR_Fig2}
%\end{figure}

\begin{figure}[h!]
\centering
\includegraphics[width=\linewidth,height=\textheight,keepaspectratio]{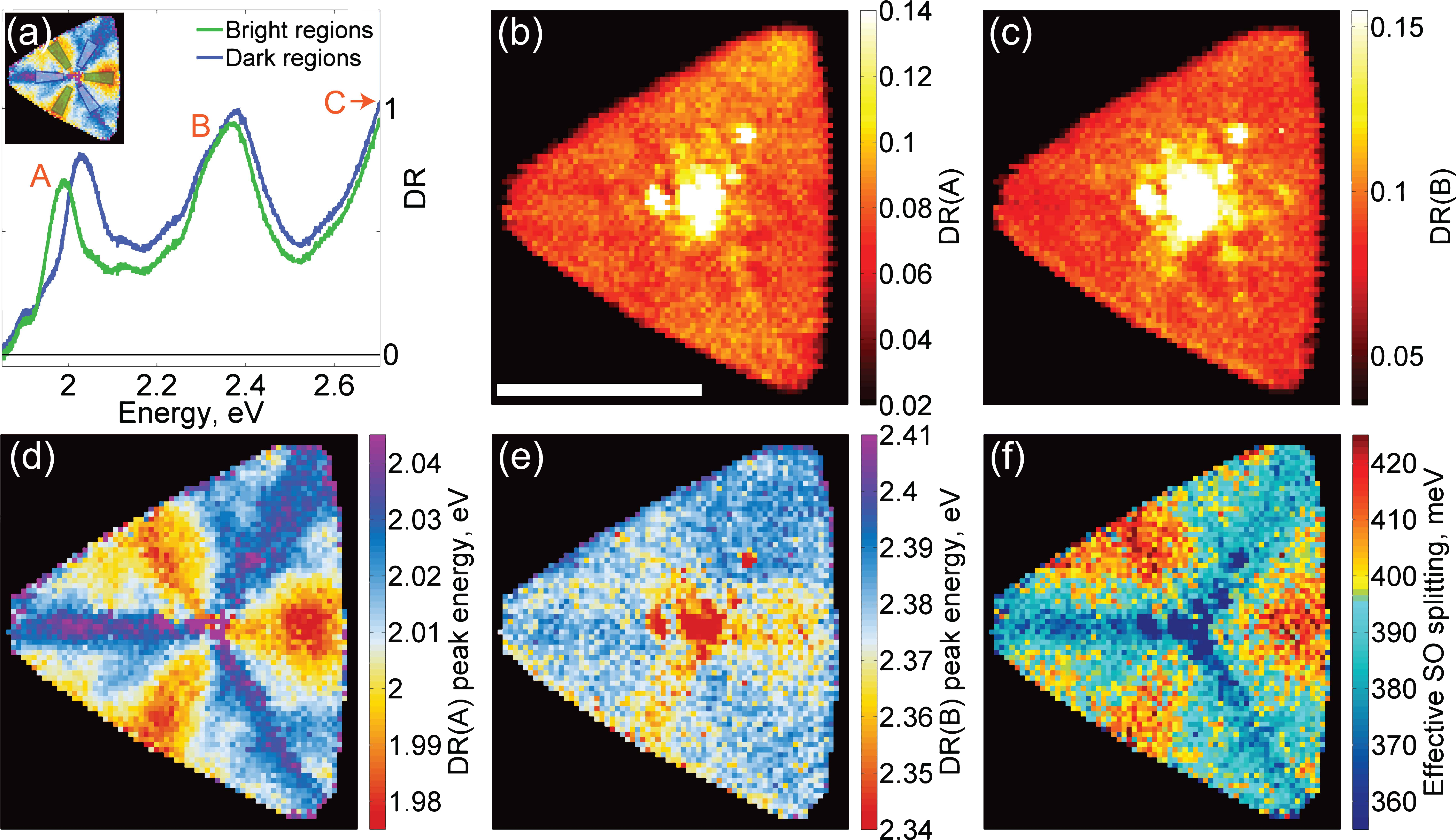}
\caption{Flake~\#1: comparison of A- and B-exciton spectral characteristics. (a) DR spectra integrated over the dark (blue) and bright (green) regions shown in inset. Spatial maps of DR peak amplitude of (b) A- and (c) B-exciton transition; (d) A- and (e) B-exciton transition energies, and (f) effective SO. The length of the scalebar in (b) corresponds to 10 $\mu$m.}
\label{fig:flake2_DR_Fig3}
\end{figure}

We obtain spatially-resolved absorption spectra of WS$_2$ monolayers (Figure~\ref{fig:flake2_DR_Fig3}) by differential reflectance (DR) hyperspectral imaging and use the approximation that DR is proportional to the absorption coefficient\cite{McIntyre1971}. This is a reasonable approximation for the case of monolayers of WS$_2$ on a thick sapphire substrate, where interference effects are minor\cite{Borys2017} and are nearly absent in the spectral range of the A-exciton absorption peak\cite{Chernikov2015}.  Figure~\ref{fig:flake2_DR_Fig3}a shows the DR spectra integrated over the same areas (shown in inset) as in the case of the corresponding PL spectra considered above. The measured DR spectra consist of the three prominent features. The A- and B-exciton transitions originate from the SO split valence and conduction bands at K symmetry points in the first Brillouin zone, whereas the onset of the third peak belongs to the C-exciton transitions in the band-nesting region\cite{Kozawa2014}. 
%****MOVED****In these measurements, it was apparent that the energy of the B-exciton transition does not vary significantly across the flake compared to that of the A-exciton transition (Figure~\ref{fig:flake2_DR_Fig2}). This trend, where the B-exciton peak energy varies to a lesser extent than the energy of A-exciton, was observed across all monolayer flakes investigated in this work (see Supporting Information). 
%Previous studies on MoS$_2$ show B-excitons to be less sensitive to the number of MoS$_2$ layers\cite{Dhakal2014,Castellanos-Gomez2016}, stacking orientation of MoS$_2$ bilayers\cite{Park2016}, position within a MoS$_2$/hBN (hexagonal boron nitride) heterostructure and whether the MoS$_2$/hBN is degraded or not\cite{Nozaki2016}. This renders B-excitons, in particular, as less sensitive to various perturbations of the intrinsic optoelectronic properties of these materials than A-excitons, although the life-time of B-excitons is shorter and corresponding spectral profile is broader.

%*******REMOVE part c (their ratio) of this figure and replace with integrated plot currently Fig2*********
% One can also distinguish fringes present across the flake in the DR amplitude maps, which are likely correlated with the substrate steps commonly present on the sapphire's (0001)-plane \cite{Zheng2015}.

The spatial maps of the A- and B-exciton absorption features (Figure~\ref{fig:flake2_DR_Fig3}) show that in contrast to PL, the absorption amplitudes of these two peaks (Figure~\ref{fig:flake2_DR_Fig3}b,c) show only minor variations across the monolayer. This suggests that the oscillator strength is not significantly changed, but rather that the changes in PL intensity are due to reduced PL QY. The bright regions near the centre of the triangles correspond to multilayers which typically give a greater reflectance contrast\cite{Dhakal2014,Castellanos-Gomez2016}. These multilayer regions are not so evident in the PL maps because the PL from multilayers is reduced and these regions lie in areas where the PL intensity is already low.
% Figure~\ref{fig:flake2_DR_Fig3}c shows the A-to-B-exciton absorption amplitudes ratio providing a better contrast on the variation of A- and B-exciton absorption amplitudes. It is clear from the map that the two amplitudes are not linearly correlated across the flake, and their ratio is larger in the dark regions suggesting different rates of absorption amplitude for A- and B-exciton transitions with the changes of overall perturbation acting on the crystal lattice. 

The A-exciton absorption peak energy (Figure~\ref{fig:flake2_DR_Fig3}d) varies significantly between the dark and bright regions, similar to the PL emission. In contrast, the B-exciton energy varies by a much smaller amount (Figure~\ref{fig:flake2_DR_Fig3}e). This is also clear in the integrated DR plots in Figure~\ref{fig:flake2_DR_Fig3}a and was observed for all flakes examined in this work (see Supplementary Material). This results in spatial variations of the effective SO splitting shown in Figure~\ref{fig:flake2_DR_Fig3}f %where the value of 27.7~meV was added to all energy differences between the A- and B-exciton peaks to take into account a constant SO splitting of the conduction band in the lattice-perturbation-free case. This value is the average SO splitting of the conduction band minimum at K symmetry points reported previously for WS$_2$ monolayers\cite{Kosmider2013,Kosmider2013a,Liu2013,Absor2016,Guo2017}. We note however, that 
which could arise due to changes in the valence band SO-splitting, the conduction band SO-splitting, variations in the Fermi energy or any combination thereof. Before addressing this question, we first consider further the properties of the A-exciton.
%*********\textit{\textcolor{red}{The effective SO splitting of the valence band maxima in the bright regions reaches $\sim$416 meV which is in agreement with the values previously derived from ARPES\cite{Latzke2015,Tanabe2016} and optical spectroscopy\cite{Zhu2015} measurements. This agreement is likely enabled by the opposite behaviour of the conduction band splitting and the valence band splitting with tensile strain\cite{Absor2016} resulting in the effective SO splitting being less affected by strain. 
%Previous first-principles calculations demonstrated an increase of the SO splitting of the valence band and the decrease of the SO splitting of the conduction band with increasing strain\cite{Absor2016}, and the larger values of the measured effective SO splitting in the bright regions support that these regions could be affected by larger strain fields.%\cite{Dery2016}
%}}***I dont think this is right. revise and maybe move to after we have determined that there is doping***
\begin{figure}[h!]
\centering
\includegraphics[width=\linewidth,height=\textheight,keepaspectratio]{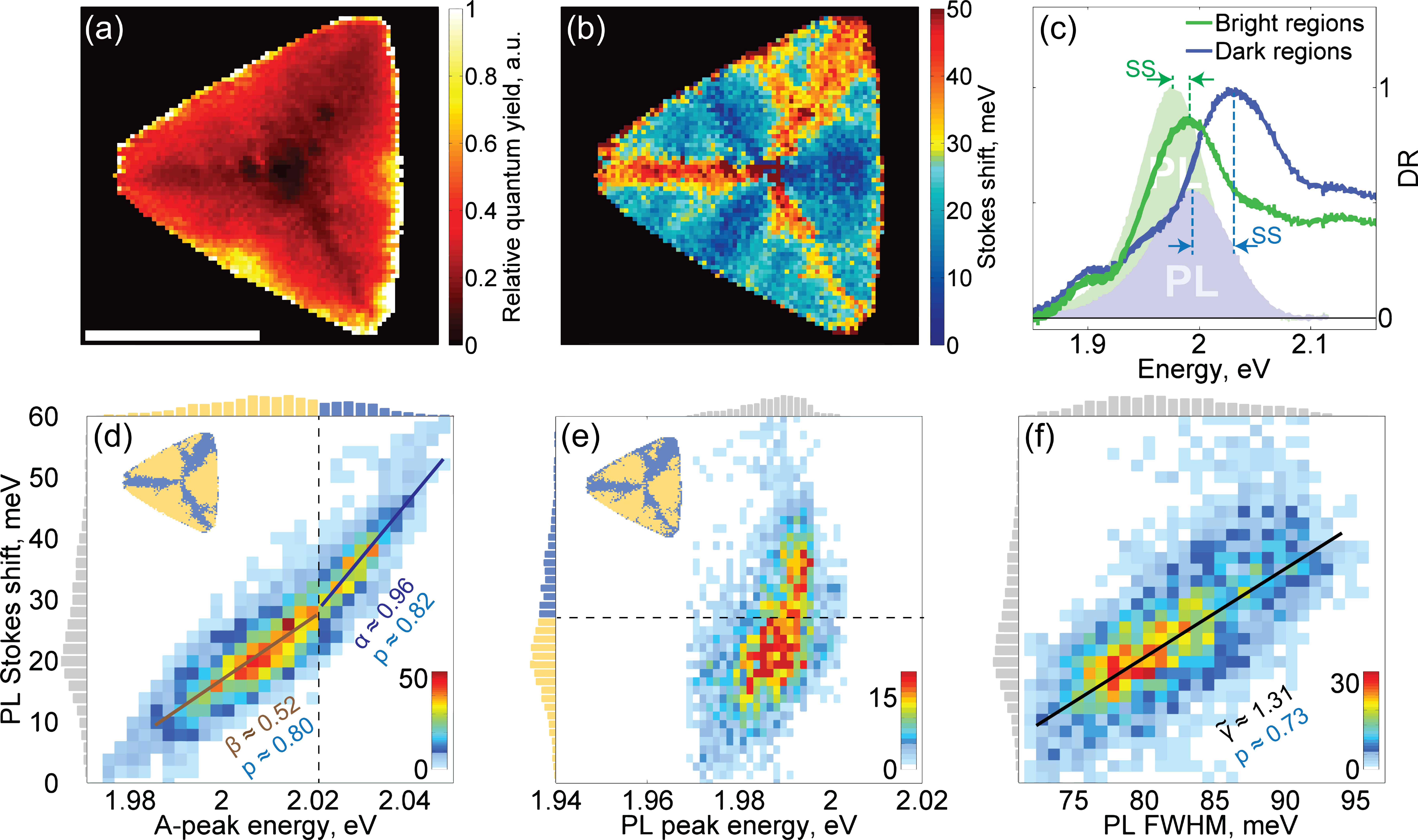}
\caption{(a) Relative PL quantum yield and (b) PL Stokes shift for the flake~\#1. (c) DR and PL spectra integrated over the areas shown in inset in Figure~\ref{fig:flake2_DR_Fig3}a. Comparison of correlation plots (2D histograms) of PL Stokes shift versus A-exciton DR peak energy (d), PL peak energy (e) and PL FWHM (f). The bin size in (d) is 0.003$\times$2.012 eV$\cdot$meV, in (e) is 0.001$\times$1.391 eV$\cdot$meV, and in (f) is 0.84$\times$1.89 meV$\cdot$meV. Horizontally and vertically integrated histograms are shown on the left and on the top of each plot in (d--f), respectively. The integrated 1D correlation plots in (d,e) were split into two parts corresponding to dark (blue) and bright (yellow) regions. In (b) it was not possible to find a vertical boundary, therefore a horizontal boundary was used instead. All boundaries were drawn from where the slope of a chosen 1D histogram changes. The data points from each side of a boundary were mapped back onto the corresponding flake (see insets in (d,e)). In (d) the two data sets corresponding to dark and bright regions were fit with linear functions. In (f) data set was fit with a single linear function. For all linear fits the extracted slopes and calculated Pearson's correlation coefficients are given. Colorbars in (d--f) reflect the number of data points within a bin. The length of the scalebar in (a) corresponds to 10 $\mu$m.}
\label{fig:flake2_PLDR_Fig4}
\end{figure}

Figure~\ref{fig:flake2_PLDR_Fig4}a,b combines the PL and DR maps discussed above to reconstruct the spatial variations of the relative PL QY and PL Stokes shift of  monolayer flake~\#1. The spatial maps of PL QY (Figure~\ref{fig:flake2_PLDR_Fig4}a) highlight that the reduction of PL intensity in the dark regions is primarily due to the lower PL QY. This could arise due to higher $n$-doping levels and/or greater defect density in the dark regions\cite{Mak2012,Mouri2013,Zande2013}. The map of PL Stokes shift (Figure~\ref{fig:flake2_PLDR_Fig4}b) shows values ranging from $\sim$5 meV in the bright regions to $\sim$55 meV in the dark regions. Integrated PL and absorption spectra in the dark and bright regions are superimposed in Figure~\ref{fig:flake2_PLDR_Fig4}c and show that the larger Stokes shift in the dark regions arises due to a large shift of the absorption peak.

\begin{figure}[h!]
\centering
\includegraphics[width=0.8\linewidth,height=\textheight,keepaspectratio]{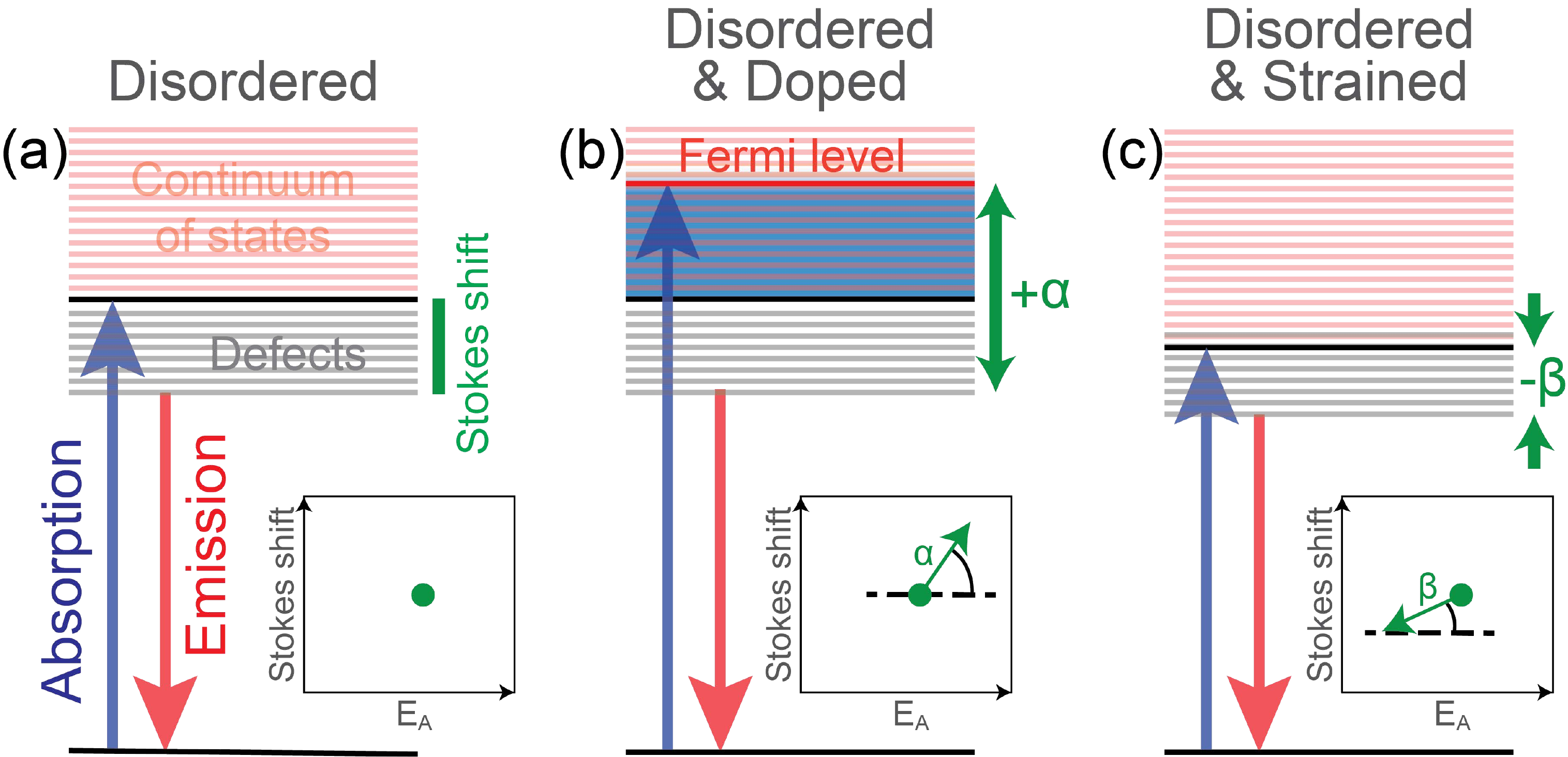}
\caption{Schematic energy level diagrams showing absorption and emission processes resulting in finite PL Stokes shifts (green). (a) The case of undoped and unstrained monolayer with disorder. (b) The case of doped monolayer with disorder; the Fermi level is denoted by the red line; free electron states are shaded in blue; PL Stokes shift increases with the rate $+\alpha$. (c) The case of strained monolayer with disorder; PL Stokes shift decreases with the rate $-\beta$; both absorption and emission exhibit redshifts. Absorption (blue arrows) and emission (red arrows) energies, defect states (light grey) and continuum of states (light red) are denoted correspondingly.}
\label{fig:effectsStrainDopingSchematic3}
\end{figure}

In general, PL Stokes shift can originate from randomly-distributed disorder potentials and interactions with optical phonons \cite{Yang1993,Valkovskii2018} (Figure~\ref{fig:effectsStrainDopingSchematic3}a), elevated electron doping density\cite{Schmitt-Rink1989,Hawrylak1991,Mak2012} (Figure~\ref{fig:effectsStrainDopingSchematic3}b), or the presence of lattice strain field\cite{Aït-Ouali1998,Niehues2018} (Figure~\ref{fig:effectsStrainDopingSchematic3}c). To be more specific, it has been shown that there is a linear correlation of the PL Stokes shift with $n$-doping\cite{Mak2012}, the amount of strain\cite{Niehues2018}, and disorder-induced PL spectral width (FWHM$_\textrm{disorder}$)\cite{Yang1993}. In the case of increased $n$-doping, the increasing Fermi energy level in the conduction band results in the A-exciton absorption peak being blue-shifted, while the PL peak energy is relatively unchanged. This occurs because the absorption process drives transitions between the top of the valence band and the Fermi edge in the conduction band, whereas the emission is from states at the bottom of the conduction band; therefore the doping contribution to PL Stokes shift varies approximately linearly with the A-exciton absorption peak energy ($E_A$). %, and can be represented as SS$_\textit{i}\sim\alpha E_A$, where $\alpha$ is the slope of a linear trend resulting from changing Fermi level in the conduction band.  
%In addition to the effects of Pauli blocking in the conduction band, the excess of free carriers due to doping can also lead to band gap renormalization (red-shift of the bottom of the conduction band) and increased screening, which can reduce the exciton binding energy. 
Tensile strain results in a red-shift of the conduction band minimum at K symmetry points leading to a red-shift of both absorption and emission peak energies with similar but not equal shift rates. The absorption energy shifts faster than the emission, leading to a Stokes shift that decreases with increasing tensile strain\cite{He2013,Niehues2018}.  %:  SS$_\textit{ii}\sim-\beta E_A$ with $\beta$ being the slope of a linear trend resulting from changing strain. 
Finally, the contribution from disorder can cause a change in the Stokes shift, with minimal change in absorption energy. The Stokes shift due to disorder will, however, be linearly correlated with the FWHM of the PL peak\cite{Yang1993}.%, and lead to a linear trend . be written as SS$_\textit{iii}\sim\gamma \textrm{FWHM}_\textrm{disorder}$, where $\gamma$ is the slope of a linear trend resulting from disorder potentials\cite{Yang1993}. Combining these considerations, in the simplest approximation the total PL Stokes shift (SS$_{\textrm{total}}$) can be represented as
%\begin{equation}
%\begin{aligned}
%\textrm{SS}_{\textrm{total}}= &\textrm{SS}_{\textit{i}}+\textrm{SS}_{\textit{ii}}+\textrm{SS}_{\textit{iii}}+g_{\textit{i},\t%extit{ii},\textit{iii}}=\\
%&\alpha A_E - \beta A_E + \gamma\textrm{FWHM}_\textrm{disorder} + g_{\textit{i},\textit{ii},\textit{iii}},
%\label{eq:Stokes}
%\end{aligned}
%\end{equation}
%where $g_{\textit{i},\textit{ii},\textit{iii}}$ reflects the possible coupling between the three contributions and also encapsulates constant $y$-intercepts of the three linear trends. Figure~\ref{fig:effectsStrainDopingSchematic3} schematically summarizes the three mentioned contributions to the total PL Stokes shift. Therefore, examination of the dependencies of PL Stokes shift on absorption energy and PL width can provide insights into the variations of strain, doping and disorder across monolayer flakes.

%To be able to discuss the three contributions to the PL Stokes shift and the observed differences between the two flakes in more details, 
To disentangle these different effects, we perform a statistical analysis correlating PL Stokes shift with A-exciton absorption peak energy (Figure~\ref{fig:flake2_PLDR_Fig4}d), PL emission peak energy (Figure~\ref{fig:flake2_PLDR_Fig4}e) and PL FWHM (Figure~\ref{fig:flake2_PLDR_Fig4}f).  Such correlation analyses have been shown previously to be powerful in separating physically distinct domains within MoS$_2$ monolayers\cite{Bao2015, Kastl2019} (using PL-Raman and PL-KPFM correlations), and disentangling the effects of strain and doping in graphene\cite{Lee2012}, MoS$_2$ monolayers\cite{Michail2016} and  graphene/MoS$_2$ heterostructures\cite{Rao2019} (using PL-Raman and Raman-Raman correlations).  %This approach to disentangling the effects of doping, strain and disorder based on the absorption and PL measurements is more intuitive than correlation of PL and Raman peaks.
The  plots shown in Fig.~\ref{fig:flake2_PLDR_Fig4}(d-f) are 2D histograms providing a basic idea of the data density while visually enhancing the clarity of trends within the overall scattered data. Each 2D histogram was integrated along the horizontal and vertical axes and reduced to corresponding 1D histograms that can be used to estimate the boundary (dotted line) separating the different spatial regions.

The correlation plot in Figure~\ref{fig:flake2_PLDR_Fig4}d reveals two distinct regions with different slopes. The region with the steeper slope, and higher absorption energy, maps onto lines extending from the centre of the monolayer flake to the apexes. In this region of the correlation plot it is clear that there is a strong correlation between the Stokes shift and absorption peak energy, which is fit with a slope $\alpha\sim0.96$ and Pearson's correlation coefficient ($p$) of 0.82. The high strength of this correlation suggests that there is one dominant origin of the changing Stokes shift in this spatial region.
%that is varying slightly along the lines from the centre to the apexes and considerably sideways.  
In contrast, for the same range of Stokes shifts there is no significant correlation with PL peak energy (Figure~\ref{fig:flake2_PLDR_Fig4}e).  This correlation of PL Stokes shift with absorption peak energy, but not with PL emission peak energy, strongly supports increased electron doping in the dark regions and, particularly, the narrower regions indicated in the insets of Figure~\ref{fig:flake2_PLDR_Fig4}, although we do not exclude the possible effects of minor compressive strain in the dark regions\cite{Absor2016,Jeong2017}.

%Governed by Eq. \ref{eq:Stokes} we fit the scattered data using linear fit models, separately in the dark and bright regions and use the Pearson correlation coefficient, $p$, to estimate the strength of the correlation (see Supporting Information). %Correlation of PL Stokes shift with the absorption and emission peak energies in the case of the flake~\#1 
%Figure~\ref{fig:hist2D_flake5_flake2_plots}a shows that in the dark regions (which correspond to the regions of highest absorption energy and largest Stokes shift) there is a strong correlation between the Stokes shift and absorption energy with a slope of $\sim$0.96 and Pearson correlation coefficient of 0.82. In contrast, for the same region, there is no significant correlation of Stokes shift with PL energy (Figure~\ref{fig:hist2D_flake5_flake2_plots}b). The strength of the correlation in Figure~\ref{fig:hist2D_flake5_flake2_plots}a suggests that there is one predominant **perturbation** that is varying considerably in this region.  The correlation of Stokes shift with absorption energy, but not with PL emission peak energy strongly supports increased doping and variations in the doping density across the dark regions.
%, in which case the emission that occurs from the charge-neutral exciton states experiences much smaller energy shifts than the absorption edge which is more sensitive to doping due to Pauli blocking of charged fermionic states (Figure~\ref{fig:effectsStrainDopingSchematic3}b).

Away from these narrow regions connecting the centre to the apexes (i.e. in the yellow regions in the inset of Figure~\ref{fig:flake2_PLDR_Fig4}d), the slope of the correlation between Stokes shift and A-exciton absorption peak energy is reduced to $\beta\sim0.52$ and the strength of the correlation is also slightly reduced, with the Pearson's correlation coefficient down to 0.8. In addition, a weak correlation between Stokes shift and PL peak energy is evident (Figure~\ref{fig:flake2_PLDR_Fig4}e).
This matches well with strain-dependent studies\cite{He2013,Niehues2018} showing that with increasing strain both absorption and emission energies red-shift, with the rate of absorption peak energy shift being larger than that of the emission peak energy  (Figure~\ref{fig:effectsStrainDopingSchematic3}c). The net result is a Stokes shift that decreases with increasing strain, alongside a decrease in absorption energy and PL peak energy, as is the case in Figure~\ref{fig:flake2_PLDR_Fig4}d,e. %In addition, PL Stokes shift is lower in the bright regions and therefore doping level there must be lower. 
%Given that this half of the correlation plot continues from the regions attributed to having higher and varying doping levels, we suggest that the doping density in the bright regions is reduced, and it is the variation of the tensile strain field that is primarily responsible for the changes in absorption energy, PL peak energy and Stokes shift in these regions. We note, however, that the Pearson's correlation coefficient is lower in this region, which suggests a more significant contribution from other sources, possibly including disorder and a small contribution from electron doping. 
We therefore suggest that in these regions, the doping density is reduced, and it is the variation of the tensile strain field that is primarily responsible for the changes in absorption energy, PL peak energy and Stokes shift in these regions. Although, we note that the Pearson's correlation coefficient is lower in this part of the correlation plots, which suggests a more significant contribution from other sources, possibly including disorder and a small contribution from electron doping. Returning to the maps of Stokes shift (Figure~\ref{fig:flake2_PLDR_Fig4}b), absorption energy (Figure~\ref{fig:flake2_DR_Fig3}d) and PL peak energy (Figure~\ref{fig:flake2_PL_Fig1}b), the changes to these spectral properties as a function of position then indicates that tensile strain increases from the lines-to-the-apexes to the centres of the sides. %This is also consistent with the variations in PL intensity and PL quantum yield.
This also explains why the two distinct slopes meet at the end of their ranges, and the trend in the correlation data varies continuously: the highly doped regions (along the lines to the apexes) occur at the regions with lowest tensile strain.

%************
Further insight into the role of disorder in the Stokes shift and overall optical properties can be gained from the correlation between the Stokes shift and the FWHM of the PL peak, as shown in Figure~\ref{fig:flake2_PLDR_Fig4}f. 
%In general,  broadening of the PL emission may result from elevated $n$-doping density \cite{Ruckenstein1987,Schmitt-Rink1989}, changes to the strain field\cite{Khatibi2018,Niehues2018}, or increased disorder.  
It has been previously reported that the contribution solely from disordered potential should result in a linear trend between the PL Stokes shift and exciton PL emission width with a definite slope of $\sim$0.55\cite{Yang1993}. The slope measured in Figure~\ref{fig:flake2_PLDR_Fig4}f is $\sim$1.31, which is much steeper (i.e. the Stokes shift increases faster than expected from just disorder induced broadening) and indicates that the PL Stokes shift in this flake is affected primarily by doping and strain. That is, the larger $n$-doping density in the dark regions broadens PL emission and the increasing tensile strain in the bright regions narrows the PL linewidth\cite{Khatibi2018,Niehues2018}.

From these measurements we can therefore make the following conclusions regarding the perturbations of this flake: along the lines from the centre to the apexes, there is an elevated (and varying) $n$-doping density. %This is consistent with previous reports that attribute elevated doping along similar narrow dark regions to a high density of grain boundaries.
Away from these lines the doping density gradually decreases and tensile strain begins to take over as the dominant factor affecting the optical properties. In this sample random defects and disorder make only minimal impact on the optical properties compared to doping and strain. 

Returning to the B-exciton, Fig.~\ref{fig:flake2_DR_Fig3}(e), the much smaller variation in absorption energy follows a similar pattern to, and correlates with, the PL peak energy. This variation is thus attributed to the changing strain across the flake. However, the absence of a substantially blue shifted region along the lines to the apex, as is seen for the A-exciton absorption (Figure~\ref{fig:flake2_DR_Fig3}d), indicates that the B-exciton energy is not significantly impacted by $n$-doping.  Intuitively, this could be due to the conduction band involved in the B-exciton transition lying above the one involving the A-exciton transition, but in WS$_2$, this is not expected to be the case\cite{Liu2013}. Alternatively, it could be due to competing influences cancelling out: for example, the predicted increase in SO splitting in the conduction band with increasing $n$-doping\cite{CB_SOsplitting} could offset the increase in Fermi energy. This remains an interesting question and the topic of further work. 

\begin{figure}[h!]
\centering
\includegraphics[width=\linewidth,height=\textheight,keepaspectratio]{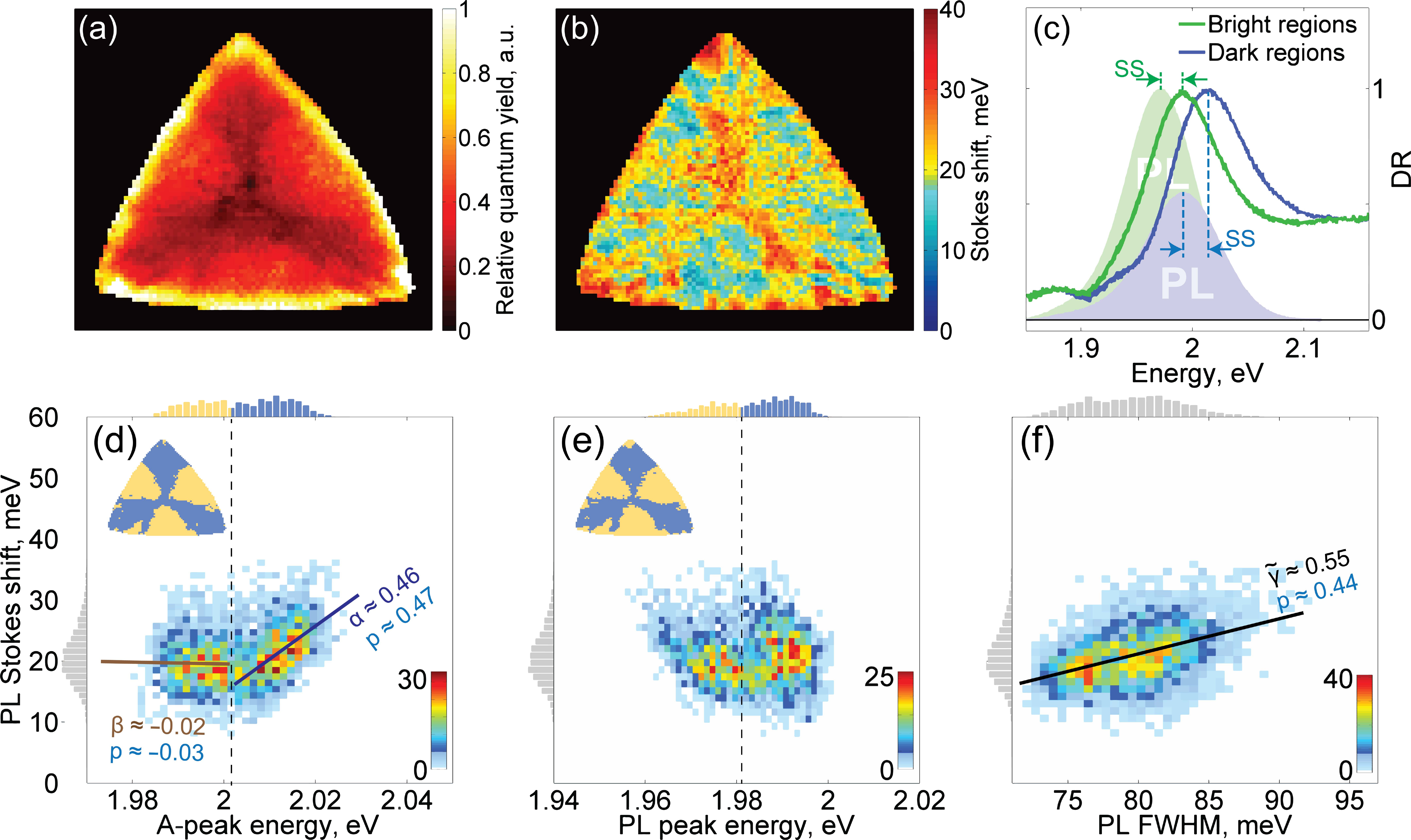}
\caption{(a) Relative PL quantum yield and (b) PL Stokes shift for the flake~\#2. (c) DR and PL spectra integrated over the areas shown in Supporting Information. Comparison of correlation plots (2D histograms) of PL Stokes shift versus (d) A-exciton DR peak energy, (e) PL peak energy and  (f) PL FWHM. The bin size in (d) is 0.02$\times$1.56 eV$\cdot$meV, in (e) is 0.02$\times$1.63 eV$\cdot$meV, and in (f) is 0.62$\times$1.31 meV$\cdot$meV. Horizontally and vertically integrated histograms are shown on the left and on the top of each plot in (d--f), respectively. The integrated 1D correlation plots in (d,e) were split into two parts corresponding to dark (blue) and bright (yellow) regions. All boundaries were drawn from where the slope of a chosen 1D histogram changes. The data points from each side of a boundary were mapped back onto the corresponding flake (see insets in (d,e)). In (d) the two data sets corresponding to dark and bright regions were fit with linear functions. In (f) data set was fit with a single linear function. For all linear fits the extracted slopes and calculated Pearson's correlation coefficients are given. Colorbars in (d--f) reflect the number of data points within a bin. The length of the scalebar in (a) corresponds to 10 $\mu$m.}
\label{fig:hist2D_flake5_flake2_plots}
\end{figure}

To further demonstrate the capability of this approach that correlates PL and absorption imaging data we applied it to several other flakes, as shown in the Supporting Information. All other flakes examined in this work show similar trends, with two different slopes in the Stokes shift versus A-exciton absorption energy plot, and with the steeper slope, corresponding to the darker regions.
The trends in the flake with behaviour furthest from the one discussed so far (flake~\#2) are shown in Figure~\ref{fig:hist2D_flake5_flake2_plots}. In this case, the variation in PL QY is similar to that for  flake~\#1, however, the PL Stokes shift only varies between $\sim$10~meV and $\sim$30~meV and does not exhibit the pronounced trigonal symmetry. (Refer to Supporting Information for PL and DR maps of flake~\#2.)
One of the differences in the preparation of the flake~\#2 was that the measurements were performed after it had been exposed to air for more than four weeks after growth, compared to less than one week for  flake~\#1. It has been shown previously that "aging" of monolayer flakes is accompanied by interactions with environmental molecules\cite{Zhang2013,Gao2016}, leading to changes in the doping density and increased disorder, while relaxation of the monolayer on the substrate can lead to reduced strain (e.g. via propagation of micro-cracks\cite{Romanov2007,Wang2015,Gao2016,Xiong2016}). The clearest sign of aging, however, is that this flake has a much thicker edge region, with brighter PL, arising from water intercalation, which occurs over time\cite{Zheng2015}.

%To be able to address the disorder contribution to the PL Stokes shift, we first consider the flake~\#2 (Figure~\ref{fig:hist2D_flake5_flake2_plots}a--f). While the distribution of PL quantum yield in this case is similar to that for the flake \#1, the PL Stokes shift only varies between $\sim$10~meV and $\sim$30~meV and does not exhibit pronounced trigonal symmetry in the corresponding spatial map. (Refer to Supporting Information for PL and DR maps of the flake~\#2.) This is most likely due to the strain relaxation occurred to different extents for different samples accompanied with the reduction of $n$-doping due to interaction with environmental molecules. 
Figure~\ref{fig:hist2D_flake5_flake2_plots}c shows integrated DR and PL spectra for flake~\#2 (see Supporting Information for integration areas for this flake) and reveals that the main difference between the two monolayers (flake~\#1 and flake~\#2) is the energy of A-exciton absorption peak in the dark regions: for the flake~\#1 the peak is significantly blue-shifted, whereas for flake~\#2 the shift is similar to the bright regions. This strongly suggests that the doping density is reduced in the dark regions of flake~\#2.
Figure~\ref{fig:hist2D_flake5_flake2_plots}d, which plots Stokes shift as a function of absorption energy, does show that there are still two distinct regions with different slopes. However, both slopes are significantly smaller, and there is a larger spread of data points (lower $p$-value). In the dark regions, the value of PL Stokes shift changes with a slope of $\sim$0.46 ($p=0.47$), almost half compared to the flake~\#1, while for the bright regions the Stokes shift is nearly constant ($p=0.03$). We attribute this to the aging process. In particular, the structural differences that were present in the flake~\#1 and likely in the flake~\#2 remain, however, over time the adsorption of other molecules from the environment reduces the overall $n$-doping and increases the amount of disorder.
%(e.g. due to the formation of micro-cracks\cite{Romanov2007,Wang2015,Gao2016,Xiong2016})
%The only contribution to the PL Stokes shift left is the contribution from disorder, which is expected to have increased in this older sample. 
The significant role of disorder is also supported by the weak anti-correlation between the PL Stokes shift and PL peak energy: lower emission peak energies correspond to a larger PL Stokes shift due to the excitons interacting with optical phonons that assist the quasiparticles in finding deeper local minima of a fluctuating local potential\cite{Yang1993,Krustok2017,Valkovskii2018}. %The observation that adjacent ends of the linear trends near the boundary (dashed line) between the dark and bright regions for both monolayers are not significantly offset (along the boundary) further supports the presence of several contributions to the PL Stokes shift acting simultaneously each to a different extent.
In addition, Figure~\ref{fig:hist2D_flake5_flake2_plots}f shows the slope of the PL Stokes shift versus PL FWHM to be $\sim$0.55. This matches the case when a disordered potential is the primary cause for a non-zero PL Stokes shift\cite{Yang1993} again supporting our conclusions that in this flake disorder induced through aging dominates the optical properties, albeit in the presence of strain and grain boundaries that were dominant in the freshly grown flake.

\section{Conclusions}

We have investigated optoelectronic properties of WS$_2$ monolayers by correlating spatially distributed emission and absorption properties. The resultant ability to measure Stokes shift and PL quantum yield allowed us to better understand the different perturbations and their effect on the optical properties. Further insights were gained from correlation plots involving the Stokes shift, allowing us to identify (in freshly grown flakes) regions where the $n$-doping density is high (and varying), and regions where tensile strain is the dominant varying perturbation. In the case of the aged flake, we were able to identify the much greater role played by disorder, likely due to the interactions between the monolayer flake and its environment. The identification of regions with high $n$-doping density was supported by the introduction of $\Delta$SM (the difference between the spectral median and the peak emission energy) as a reliable fitting-free estimation of the relative charging energy of trions, which varies with the doping density. In contrast, these measurements also revealed that the B-exciton energy is apparently unaffected by increased $n$-doping. While we discussed several possibilities further work is needed to understand the origin of this surprising effect.

%, in contrast to the previous observations for MoS$_2$ monolayers\cite{Mak2012,Dhakal2014,Borys2017}, 
% Bexcitons in WS$_2$ monolayers demonstrate less sensitivity to the variations of the overall perturbations acting upon the crystal structure. More work is required to understand these differences in the behaviour of B-excitons between different TMdC monolayers. 
%Finally, we suggest that the dendritic patterns observed in this work may have originated from the crystal growth being initiated by multiple nucleation centres of a complex shape resulting in the competition between several growth processes.
Finally, the correlation plots that helped disentangle the different perturbations here represent just a small subset of the large multi-dimensional data-cube that is formed from the spectral parameters obtained from the measured data. Development of other approaches to better breakdown this vast dataset may lead to further insights and even better capabilities to identify different types of perturbation in 2D materials.

\section{Experimental}

\subsection{Sample preparation}

The sample preparation was performed in a similar way as described in Ref.~[\cite{Zhang2018}].

\subsection{Photoluminescence measurements}

The spatially-resolved PL measurements 
were performed using the frames of an inverted microscope (Nikon Eclipse T\textit{i}-U). Linearly polarized coherent excitation was provided by a cw laser diode (Thorlabs, L405P20, mounted into the TE-Cooled Mount TCLDM9, operated by a temperature controller TED200C and a laser diode controller LDC205C) tuned to the wavelength of $\sim$410 nm. The radiation is sent to a dichroic mirror reflecting wavelengths below and transmitting wavelengths above $\sim$593 nm. After the dichroic mirror, the laser is focused onto the sample by a 100x objective lens (Olympus) with numerical aperture NA=0.95. The power density at the sample was estimated to be $\sim$2.8 MW/cm$^2$. The induced PL from the samples is collected by the same objective lens (\textit{epi}-fluorescence geometry) and is transmitted through the dichroic mirror and a lens (200 mm focal length). At the image plane of the lens, a pinhole of 150 $\mu$m diameter transmits the sample's region with $\sim$750 nm diameter. The signal transmitted through the pinhole is coupled to a multi-mode fiber (Avantes, 200 $\mu$m core diameter, 2 m length) by a free-space fiber coupler (Thorlabs, F810SMA-635). The fiber is attached to a spectrometer (Avantes, Avaspec-2048), and a 400 ms integration time was used at each point.

The sample was mounted on a dual-channel XY motorized translation stage (Applied Scientific Instrumentation, MS-2000), allowing for raster scanning mode. The scanning step was chosen to be 250 nm (flakes \#1,2,5,6,7) and 500 nm (flakes \#3,4). The communication between the motorized stage and the spectrometer was established by means of a LabView code. All measurements have been performed at room temperature. The estimated spatial resolution of the setup is $\sim$300 nm (see Supporting Information).

\subsection{Differential reflectance measurements}

The spatially-resolved DR measurements were performed using the same PL setup with only slight differences. Instead of the diode laser as an excitation source, a spectrally broad ($\sim$400--800 nm) incoherent white light (CoolLED, \textit{p}E-100) was directed to the sample via a 50:50 beam splitter. The brightness of white light allowed to use a pinhole of 100 $\mu$m diameter at the image plane transmitting a region of $\sim$500 nm diameter. The integration time was 150 ms. The resolution of the technique was estimated to be $\sim$380 nm at $\sim$620 nm excitation wavelength (see Supporting Information).

\begin{acknowledgement}

This work was supported by the Australian Research Council Centre of Excellence for Future Low-Energy Electronics Technologies (CE170100039).

\end{acknowledgement}

\begin{suppinfo}

%A listing of the contents of each file supplied as Supporting Information
%should be included. For instructions on what should be included in the
%Supporting Information as well as how to prepare this material for
%publications, refer to the journal's Instructions for Authors.

The following files are available free of charge.
\begin{itemize}
  \item Filename: acsnano\_suppl\_PK.pdf
  \item Filename: PL\_SS\_anim.avi
\end{itemize}

\end{suppinfo}

\bibliography{achemso-demo}

\end{document}